\documentclass[aps,prb,twocolumn,superscriptaddress,groupedaddress,citeautoscript,preprintnumbers,floatfix]{revtex4}
\pdfoutput=1
\usepackage[latin9]{inputenc}
\usepackage[english]{babel}
\usepackage{graphicx}
\usepackage{bm}
\usepackage{dcolumn} 
\usepackage{amsfonts}
\usepackage{natbib}
\usepackage{upgreek}
\usepackage{amssymb}
\usepackage{amsmath}
\usepackage{physics}
\usepackage{graphicx}
\usepackage{lipsum}
\usepackage{float}
\usepackage{doi}
\usepackage{url}
\usepackage{hyperref}
\hypersetup{
    colorlinks=true,
    urlcolor=blue,
    linkcolor=blue,
    citecolor=blue}
\usepackage[dvipsnames]{xcolor}
\graphicspath{
{./Figures/}
}

\newcommand*{\citen}[1]{%
  \begingroup
 \romannumeral-`\x 
    \setcitestyle{numbers}%
    \cite{#1}%
  \endgroup   
}

\newcommand*\diff{\mathop{}\!\mathrm{d}}

\newcommand{\di}{{i\mkern1mu}}

\begin{document}

\title{Steady one-dimensional domain wall motion in biaxial ferromagnets: mapping of the Landau-Lifshitz equation to the sine-Gordon equation}

\author{R. Rama-Eiroa}
\email{ricardo.rama@ehu.eus}
\affiliation{Donostia International Physics Center, 20018 San Sebasti\'an, Spain}
\affiliation{Materials Physics Department, University of the Basque Country, UPV/EHU, 20018 San Sebasti\'an, Spain}
\author{R. M. Otxoa}
\affiliation{Hitachi Cambridge Laboratory, J. J. Thomson Avenue, Cambridge CB3 0HE, United Kingdom}
\affiliation{Donostia International Physics Center, 20018 San Sebasti\'an, Spain}
\author{P. E. Roy}
\affiliation{Hitachi Cambridge Laboratory, J. J. Thomson Avenue, Cambridge CB3 0HE, United Kingdom}
\author{K. Y. Guslienko}
\email{kostyantyn.gusliyenko@ehu.eus}
\affiliation{Materials Physics Department, University of the Basque Country, UPV/EHU, 20018 San Sebasti\'an, Spain}
\email{kostyantyn.gusliyenko@ehu.eus}
\affiliation{IKERBASQUE, the Basque Foundation for Science, 48013 Bilbao, Spain}

\date{\today}

\begin{abstract}
Motivated by the difference between the dynamics of magnetization textures in ferromagnets and antiferromagnets, the Landau-Lifshitz equation of motion is explored. A typical one-dimensional domain wall in a bulk ferromagnet with biaxial magnetic anisotropy is considered. In the framework of Walker-type of solutions of steady-state ferromagnetic domain wall motion, the reduction of the non-linear Landau-Lifshitz equation to a Lorentz-invariant sine-Gordon equation typical for antiferromagnets is formally possible for velocities lower than a critical velocity of the topological soliton. The velocity dependence of the domain wall energy and the domain wall width are expressed in the relativistic-like form in the limit of large ratio of the easy-plane/easy-axis anisotropy constants. It is shown that the mapping of the Landau-Lifshitz equation of motion to the sine-Gordon equation can be performed only by going beyond the steady-motion Walker-type of solutions.
\end{abstract}

\maketitle

\section{Introduction}\label{sec:intro}

The exchange integral in the Heisenberg Hamiltonian determines the relative orientation between neighboring spins. When it is positive, the favored magnetization orientation between neighboring atomic sites is parallel, being this type of media known as ferromagnets (FM), while when it is negative, an antiparallel orientation is preferred, being called antiferromagnets (AFM). Both types of systems have similar types of spin excitations, such as spin waves (SW) and domain walls (DW), and the magnetization dynamics can be described, in continuum field approximation, by the Landau-Lifshitz (LL) equation of motion. \cite{LandauLifshitz1935} Interestingly, the dynamics in FM and AFM results in different SW frequency modes with a natural frequency of the order of GHz and THz, respectively. This is due to the typical order of magnitude of the parameters that control, in each case,  the frequencies of SW excitations, relying on the magnetic anisotropy field in FM and on the intersublattice exchange field in the case of AFM. \cite{GomonayBaltzBrataasEtAl2018} Likewise, the dynamics of magnetic textures presents not only quantitative differences between FM and AFM, but also qualitative ones. \cite{KosevichIvanovKovalev1990} The stable DW dynamics in FM is possible up to a limiting velocity, from which intrinsic instabilities appear in the propagating magnetic texture due to the combination of internal translational and oscillatory modes, which is known as Walker breakdown (WB). \cite{SchryerWalker1974} On the other hand, in AFM it is possible to reach higher velocities in a stable steady-state-like motion existing, however, a limit that cannot be exceeded, which is given by the maximum magnon group velocity in the medium. This is because the DW dynamics in AFM can be described, in the framework of the non-linear $\sigma$-model, through a Lorentz-invariant relativistic-like expression known as sine-Gordon (SG) equation. \cite{Fradkin2013} As a consequence, while the dynamics of DW in FM is described in the terms of conventional Galilean dynamics, in the case of AFM the movement of the aforementioned magnetic textures will follow the precepts of special relativity. All of this is due to a single change in the Heisenberg Hamiltonian. However, the dynamics of a DW in FM and AFM is indistinguishable for weak effective fields below the threshold associated with the WB, provided its evolution obeys a steady-state-like regime. The fact that sizeable magnetization deviations with respect to the DW configuration at rest do not arise in AFM is due to the rigidity conferred to the magnetic texture in AFM due to the strong exchange interaction between neighboring spins.

The magnetization switching in FM and AFM is typically related to the nucleation and propagation of an inhomogeneous magnetization reversal mode. To implement spintronic devices whose functionality is based on the propagation of inhomogeneous magnetization textures, ultrafast and controllable dynamics is essential in order to reduce switching time. For FM, the fundamental problem in this context lies in the difficulty of reaching high speeds for the reversal mode propagation while preserving stability. Thus, AFM have been erected as a solid alternative, at least theoretically, because DW in these media can reach speeds of the order of tens of km/s without entering an irregular regime. \cite{shiino2016antiferromagnetic} However, the usefulness of AFM in the field of spintronics has been rather directed so far to a passive role, such as a necessary element to convert a FM free layer into a pinning one through the exchange field bias generated by it. This has been mainly due to the difficulty of exciting and tracking the dynamics of magnetic textures in this type of systems, as opposed to FM, at least until very recently in a very particular type of structures. \cite{gomonay2016high} Accordingly, many efforts have been invested in trying to obtain higher velocities in FM while ensuring the integrity of magnetic textures. This could be possible if the appearance of instabilities induced by the WB could be avoided, or at least delayed. Some results obtained through micromagnetic simulations show that, in fact, this is possible. In this direction, it has been observed that it is possible to eradicate the WB for one-dimensional (1D) DW in linear magnetic chains, \cite{wieser2010domain} and also for two-dimensional (2D) DW in FM nanowires. \cite{YanHertelPRL2010} However, the speed of the magnetic texture will be limited in this case by the minimum phase velocity of the SW of the medium. If this threshold is exceeded, the DW will begin to emit SW, \cite{wieser2010domain} which is known as the spin Cherenkov effect, \cite{YanAndreasKakayEtAl2011} and is the kind of phenomenon that is theoretically foreseen in AFM too.  Also, in the context of one-dimensional DW, the inclusion of a Dzyaloshinskii-Moriya exchange interaction in ultrathin films with perpendicular anisotropy results in the stabilization of the magnetic texture, making it possible to delay the appearance of the instabilities and to increase the maximum DW velocity before this phenomenon begins. \cite{ThiavilleRohartJueEtAl2012} However, these approaches present challenges from an experimental point of view for their implementation.

Given this background, an alternative would be to consider analytically the dynamics of magnetic textures in a FM system as simple as possible and try to reduce the LL equation to a SG-like expression as in AFM. If it is  possible to find a situation in which this happened, perhaps its experimental implementation could be addressed, and even more complicated systems could be considered. The 1D motion of a DW is the simplest case, in which the magnetization configuration can be considered as a function of only one spatial coordinate. The next step would be to reduce the non-linear LL equation into a simpler non-linear expression. For this, there are two main approaches: i) the method of collective coordinates, and ii) the asymptotic method. The first method is based on the inclusion of the DW center position and the azimuthal angle of the DW magnetization as the generalized coordinates of the system. This allows, ultimately, to reduce the LL equation in a system of coupled differential equations. \cite{SchryerWalker1974, ThiavilleRohartJueEtAl2012} The second method aims to describe the dynamics of the magnetic texture in terms of a dimensionless parameter that allows to apply perturbation theory when it can be considered small. In this context, this condition can be transferred to one of the angles that describe the magnetization. \cite{KosevichIvanovKovalev1990} Our work will be framed within this second approach.

One of the simplest systems that can be evaluated analytically is a FM system with biaxial anisotropy in which the dynamics of a DW is 1D, if the dynamics of a magnetic texture wants to be addressed. In this line, there is, in fact, a simpler situation, in which a 1D FM that supports a hard-axis anisotropy whose induced symmetry is broken by an in-plane magnetic field is considered, which allows to perform the aforementioned mapping, giving rise to a kink solution not compatible with the presence of DW. \cite{mikeska1981solitons} However, as mentioned, the interesting thing would be to evaluate the dynamics of DW with potential applicability. Within the Walker approximation, where solutions to dynamic equations are sought under the assumption that the azimuthal angle is independent of the spatial coordinate, \cite{SchryerWalker1974} it is possible, in fact, to find an exact solution for the aforementioned system, which was demonstrated by Schl\"{o}mann. \cite{Schloemann1971} In the more realistic context in which one works beyond the Walker-type of solutions, it was heuristically demonstrated by Enz that it is possible to reduce the LL equation into a SG-like expression. \cite{Enz1964, DoddMorrisEilbeckEtAl1982} However, this type of approach contradicts the steady-state DW motion regime in an infinite medium, where the spatial and temporal derivatives of magnetization cannot be considered as independent. Also, within the context of the mapping proposed by Enz, it has not been found yet what is the expression for the DW energy, or if this solution is stable or not. If it were the case that it was stable, and that the associated DW energy is lower than that in the case of steady-state Walker-type of solutions, it could be confirmed that there is, in fact, a case in which the dynamics of a DW in a biaxial FM can be described by SG-like expression for at least a restricted range of velocities for which this solution tends to the exact Schl\"{o}mann solution.

Therefore, in this article we consider a 1D DW in a bulk FM with biaxial magnetic anisotropy. The underlying physical basic principles of the 1D magnetic soliton theory are presented in Sec. \ref{sec:model}. The approach introduced by Schl\"{o}mann in which the DW dynamics is parameterized, without dissipation, through the dispersion relation of the linear SW with complex wave vector and frequency, which reside in the tails of the moving soliton, \cite{Schloemann1971} is introduced in Sec. \ref{section:limit}.  It is shown that in the case of a biaxial magnetic anisotropy of the easy-plane/easy-axis type, the maximum speed of the steady DW motion cannot exceed the maximum phase velocity of the linear SW with imaginary wave vector for the Walker-type of solutions. In Sec. \ref{section:gordon} we consider the mapping of the LL equation of motion to the more simple SG equation within the Walker approximation and show that the mapping  can be performed only by going beyond the steady-motion Walker-type of solutions assuming a constant magnetization azimuthal angle. In order to corroborate that in the case of large easy-plane anisotropy the DW dynamics obeys the precepts of special relativity, this situation was explored using atomistic spin dynamics simulations, which is exposed in Sec. \ref{section:simulations}. Finally, conclusions are set out in Sec. \ref{section:conclusions}. 

\section{Theoretical basis}\label{sec:model}

\begin{figure}[!ht]
\includegraphics[width=8cm]{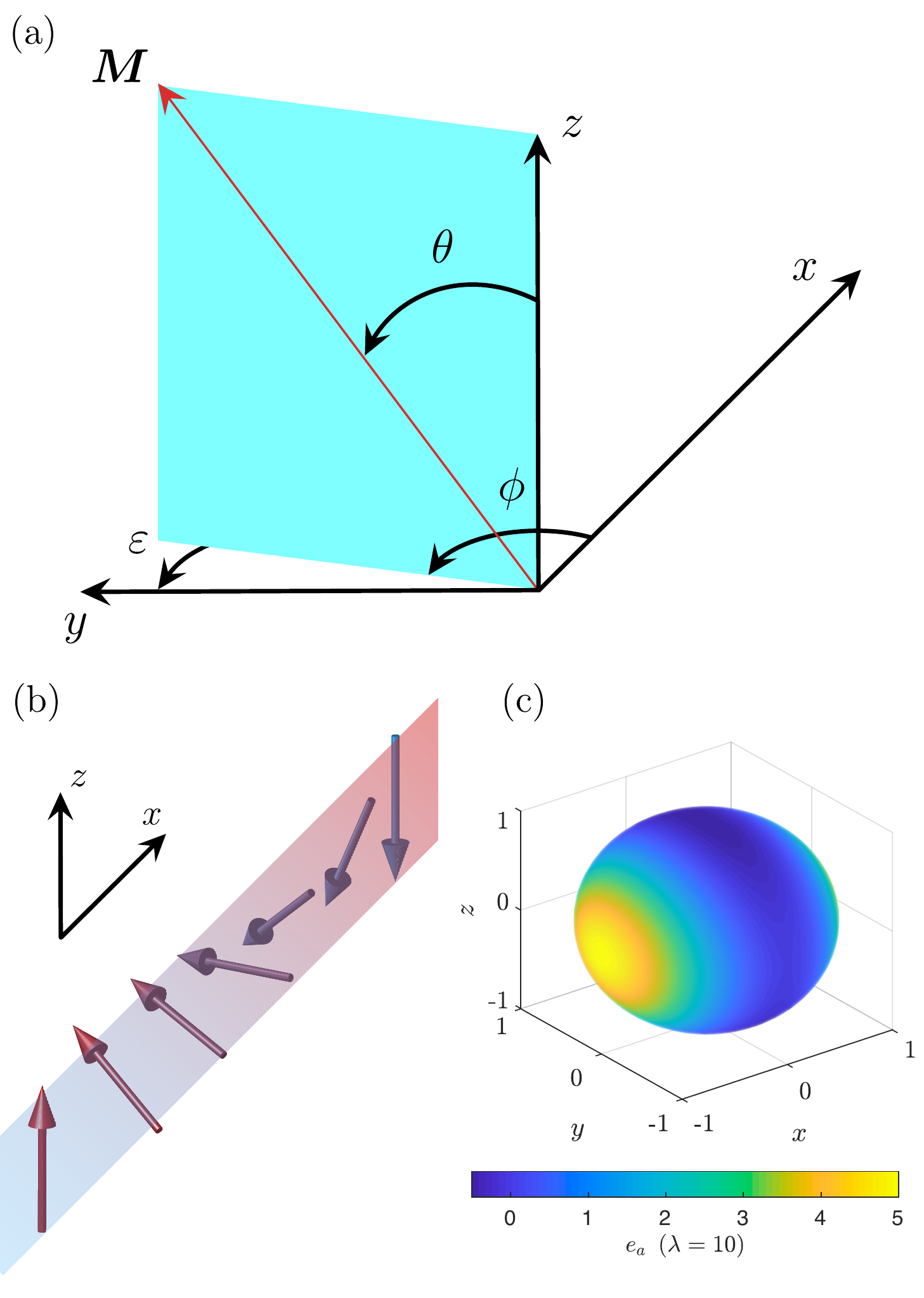}
\caption{1D DW magnetization configuration. (a) Definition of the magnetization vector $\boldsymbol{M}$ in terms of the polar $\theta$ and azimuthal $\phi$ angles relative to a Cartesian coordinate system. The angle $\varepsilon = \pi /2- \phi$ describes deviation of the magnetization from the static $yz$ DW plane. (b) Sketch of the DW magnetization configuration along the $ x $-{\it th} direction of motion. (c) Spatial distribution of the anisotropy energy density $ e_a \left( \boldsymbol{m} \right) $, in $2K_z$ units, for $ \lambda = 10 $.} 
\label{im:1}
\end{figure}

We consider a 1D DW in a bulk anisotropic FM, as sketched in Fig. \ref{im:1}. The Bloch-type DW at rest is located in the $yz$ easy-plane (see Fig. \ref{im:1} (a)), and moves along the $x$-{\it th} direction (see Fig. \ref{im:1} (b)). The total magnetic energy of the system per unit DW square (per unit area) is $E \left[ \boldsymbol{m} \right]= \int \diff x \, e \left( \boldsymbol{m} \right)$, being $e$ the energy density which, in continuum approximation, is given by $e \left( \boldsymbol{m} \right)=A \, {\left( \partial_x \boldsymbol{m} \right)}^2+e_a \left( \boldsymbol{m} \right) + e_m \left( \boldsymbol{m} \right)$. Here, $A$ represents the exchange stiffness constant, $\boldsymbol{m} \left( x, t \right) = \boldsymbol{M} \left( x, t \right) / M_s$ denotes the unit magnetization vector, $M_s$ is the saturation magnetization, $e_a$ is  the anisotropy energy density, and $e_m$ stands for the magnetostatic energy density. We consider a general quadratic form for $ e_a \left( \boldsymbol{m} \right)$ that, accounting for the restriction ${\boldsymbol{m}}^2=1$, can be expressed in the form of a biaxial anisotropy, $e_a \left( \boldsymbol{m} \right) = K_x m^2_x-K_z m^2_z$ (see Fig. \ref{im:1} (c) for the particular case $\lambda=10$, where  $ \lambda = K_x / K_z $, being the anisotropy energy in $2 K_z$ units). Assuming that $ K_x, K_z> 0 $, it is possible to define the uniform ``vacuum" state of magnetization far from the DW center, $ x \rightarrow \pm \infty $, and that the anisotropy $K_x$ is of an  easy-plane type. The magnetostatic energy density for the case of a bulk FM with the magnetization varying along the $x$-{\it th} direction is local, $e_m \left( \boldsymbol{m} \right)= 2 \pi M^2_s m^2_x$, and results in the renormalization of the anisotropy constant $ K_x $. The same expression for $e_a \left( \boldsymbol{m} \right)$ can be applied for thin magnetic stripes with  $e_m \left( \boldsymbol{m} \right)= 2 \pi M^2_s m^2_z$ (being in this case absorbed by $ K_z $), when the $y$-{\it th} component of the demagnetizing field is neglected. \cite{ThiavilleRohartJueEtAl2012} Another possibility is to define that the DW could propagate along the $z$-{\it th} direction, which belongs to the $yz$ easy-plane, there being, in this case, an easy-axis direction that breaks the symmetry imposed by the hard-axis anisotropy, either along the $y$-{\it th} or $z$-{\it th} direction. This would constitute a situation similar to that described above, with the difference that a N\'eel-type DW would be favored in this case.

We parameterize the unit magnetization vector using the spherical angles, $ \boldsymbol{m} =\boldsymbol{m} \left( \theta, \phi \right)$. The angles $ \theta, \phi $ are functions of the spatial coordinate $ x $ and time $ t $. Writing the DW energy density in units of $ 2K_z $ and lengths in units of the static DW width $ \Delta_0 = \sqrt{A / K_z} $, an expression for the energy that depends on a single dimensionless parameter $ \lambda $ can be written
\begin{equation}
e \left( \theta , \phi \right) = \frac{1}{2} \left[ {\left( \theta_x \right)}^2 + \left( 1+ \lambda \cos^2  \phi  +{\left( \phi_x \right)}^2 \right)  \sin^2 \theta \right],
\label{eq:1}
\end{equation}
where the spatial variable subscript $x$ indicates derivative with respect to it. The corresponding effective Lagrangian density for FM is given by $\mathcal{L}\left( \theta, \phi \right) = e \left( \theta, \phi \right) +\dot{\phi} \cos \theta $, \cite{Doering1948} where overdot means derivative with respect to time. Henceforth time is expressed in units of $ t_0=1/ \gamma H_a $, where $ \gamma $ is the gyromagnetic ratio, and $ H_a = 2 K_z / M_s $. The LL equations of motion in the angular representation, $\dot{\theta} \, \sin \theta=-\delta e / \delta \phi$ and $\dot{\phi} \, \sin  \theta= \delta e / \delta \theta$, can be found from a first variation of the Lagrangian taking into account the energy density given by Eq. \eqref{eq:1}, which results in
\begin{equation}
\begin{aligned}
\dot{\theta} \, \sin \theta= \left[ \lambda \cos  \phi  \sin  \phi +\phi_{xx} \right] \sin^2  \theta + \theta_x \, \phi_x \sin  2 \theta , \, \\
\dot{\phi} \, \sin \theta=\left[ 1+\lambda \cos^2  \phi  + {\left( \phi_x \right)}^2 \right]  \cos  \theta  \sin  \theta -\theta_{xx}.
\end{aligned}
\label{eq:2}
\end{equation}

The system of Eqs. \eqref{eq:2} has been intensively investigated in literature along with its integrals of motion. \cite{KosevichIvanovKovalev1990} For the particular case of a moving DW, the focus is usually on the steady-state motion Walker-type of solutions assuming $\phi_x = 0$. \cite{SchryerWalker1974} We will work within the framework of the main assumption of the theory of 1D topological magnetic solitons, \cite{KosevichIvanovKovalev1990} which is known as the ``traveling wave" {\it ansatz}, i.e., that the solutions of Eqs. \eqref{eq:2} can be written in the form $\theta= \theta \left( \xi \right)$, $\phi= \tilde{\omega} t+ \phi_0 \left( \xi \right)$, where $\xi=x-vt$, being $v$ the soliton velocity, and $\tilde{\omega}$ the soliton precession frequency in the moving frame with velocity $v$. The  moving soliton is treated as a bounded state of many SW (magnons), and the steady velocity of the soliton is interpreted as the group velocity of the SW packet, $v=v_g$. The frequencies in the laboratory frame, $\omega$, and moving frame, $\tilde{\omega}$, are related by $\tilde{\omega}=\omega - \boldsymbol{k} \cdot \boldsymbol{v}_g$. Here, $\boldsymbol{v}_g= \partial \omega / \partial \boldsymbol{k}$ denotes the group velocity of the linear SW, and $\boldsymbol{k}= k \boldsymbol{\hat{x}}$. Therefore, it is natural that $\omega=\tilde{\omega}+vk$. Those magnetic solitons with $\tilde{\omega} \neq 0$ are known as precession or dynamic solitons. \cite{KosevichIvanovKovalev1990} 

\section{Moving domain wall energy and critical velocities}\label{section:limit}

The current approach assumes that the calculation of the DW energy and the limiting DW velocities is done through the spectra of the linear SW that reside in the DW tails in the case of a saturated FM. Far from the center of the moving DW, the magnetization can be considered as uniform and parallel to the $z$-{\it th} anisotropy easy-axis, see Fig. \ref{im:1} (b). The magnetization dynamics outside the DW can be described in terms of small amplitude (linear) SW assuming a complex wave vector and frequency. \cite{Schloemann1971} Therefore, the DW dynamics can be described considering the SW of its tails as long as the DW magnetization configuration does not change. The linear SW dispersion relation for a biaxial FM, $\omega \left( k \right) $, is well known, and can be deduced from the linearization of Eqs. \eqref{eq:2} with respect to the ground state at the tails of the DW. It is explicitly given by the dispersion equation $\omega^2= \left( 1+k^2 \right) \left( 1+ \lambda + k^2 \right)$. The generalization to the complex wave numbers and frequencies is straightforward
\begin{equation}
\Omega^2=\left( 1+K^2 \right) \left( 1+ \lambda + K^2 \right),
\label{eq:3}
\end{equation}
where $\Omega=\omega+\di \kappa v$, and $K=k+\di \kappa$.

From now on, we only consider stationary soliton motion, which assume $\tilde{\omega}=0$. This allows to write the expression $\Omega = v K$. One can find from  Eq. \eqref{eq:3} that the velocity is real in two regions disconnected from each other, $ \left[ 0, v_- \right] $ and $ \left[ \left. v_+, \infty \right) \right. $, where $ v_{\pm} = \sqrt{1+ \lambda} \pm 1 $.\cite{Schloemann1971, EleonskiKirovaKulagin1978} We note that velocities are in units of $\Delta_0 / t_0 = 2 \gamma \sqrt{ A K_z}/M_s$. The first critical velocity, $v_-$, possesses physical sense of the maximum phase velocity of SW with imaginary wave vector $K= \di \kappa$ and imaginary frequency $\Omega= \di \kappa v$. In fact, this velocity corresponds to the maximum DW velocity for the case of the steady-state motion regime. \cite{SobolevHuangChen1995} The velocity $v_-$ is higher than the critical Walker velocity, $v_{\textrm{W}}$, in a uniaxial FM assuming a driving force due to an external magnetic field or a spin polarized current. However, the ratio $v_-$/$v_{\textrm{W}}$  is not very large,  $v_-$/$v_{\textrm{W}}$ =$ \sqrt{2 \, (2+ \lambda)} \, (\sqrt{1+ \lambda} - 1)/ \lambda$, and approaches $ \sqrt{2}$ at $\lambda \gg 1$.  \cite{SchryerWalker1974} The second critical velocity, $v_+$, can be interpreted as the minimal phase velocity of SW with real frequency $\Omega = \omega$ and real wave vector $K=k$. The steady motion of DW (topological solitons satisfying the boundary conditions $\theta \left( \pm \infty \right) = 0,\pi$) is possible only within the interval $\left[ 0, v_- \right]$. These DW solutions satisfy the condition 
$\phi\left( \xi \right) = \mathrm{const}$, and describe Bloch DW ($\phi = \pm \pi /2 $), N\'eel DW ($ \phi =0, \pi$), or a hybrid DW (other values of $\phi$). The complicated solutions existing within the velocity interval $\left( v_-, v_+ \right)$ accounts for solitary magnetization waves, non-topological solitons, satisfying the boundary condition $\theta \left( \pm \infty \right) =0$. The region $ \left[ \left. v_+, \infty \right) \right. $  accommodates non-linear SW with real wave vector $k$. \cite{EleonskiKirovaKulagin1978}

It is possible to find an explicit form of the complex wave vector dependence on the soliton velocity, $ K \left (v \right) $, from Eq. \eqref{eq:3}, which is given by
\begin{equation}
K^2 \left( v \right) = \frac{1}{2} \left( v^2 - \lambda \right)-1 \pm \frac{1}{2} \sqrt{\left( v^2-v^2_- \right) \left( v^2-v^2_+  \right)} \, .
\label{eq:4}
\end{equation}

The minus sign in Eq. \eqref{eq:4} corresponds to an unstable DW solution at $v < v_{-}$ (N\'eel DW at $v=0$). \cite{Schloemann1971} The plus sign in Eq. \eqref{eq:4} holds for a stable solution (Bloch DW at $v=0$). \cite{MagyariThomas1985}

The first integral of Eqs. \eqref{eq:2}, ${\left( \theta_x \right)}^2+{\left( \phi_x \right)}^2 \, \sin^2 \theta=\sin^2 \theta \left[ 1+ \lambda \cos^2 \phi \right]$, \cite{EleonskiiKirovaKulagin1976} allows to calculate the energy, $E_{\mathrm{DW}}$, for the stable solution as the doubled exchange energy. The DW energy for the Walker-type of solutions (that is, $\phi = \mathrm{const}$), $E_{\mathrm{DW}} \left( v \right) = E_0 \, \kappa \left( v \right)$ in units of $2K_z \Delta_0$, increases with velocity up to a maximal value $E_{\mathrm{DW}}=E_0{\left( 1+\lambda \right)}^{1/4}$, where $E_0=2$ is the static DW energy. The DW width, $\Delta \left( v \right) = 1/ \kappa \left( v \right)$, decreases with velocity (dynamical contraction) reaching the finite minimal value $\Delta \left( v_- \right) = {\left( 1+\lambda \right)}^{-1/4}$. In addition, it is possible to verify that in the $\left[ 0, v_- \right]$ region the DW plane orientation angle, $\phi \left( v \right)$, decreases as the velocity $v$ increases from $\pi /2$ to $\phi \left( v_- \right) = \arccos \sqrt{v_- / \lambda } \,$. The decomposition of $E_{\mathrm{DW}}\left( v \right)$ in series on small velocities ($v \ll v_-$), $E_{\mathrm{DW}}=E_0+m_{\mathrm{DW}}v^2/2$, allows to find the DW D{\"o}ring mass for the stable Bloch-like DW (in absolute units) $m_{\mathrm{DW}}=1/ 2 \pi \gamma^2 \Delta_0$. \cite{Doering1948, HubertSchaefer2008} Therefore, Eqs. (\ref{eq:2},~\ref{eq:3}) lead to correct results for relatively small SW velocities lying within the interval $\left[ 0, v_- \right]$. The second, unstable, solution of Eq. \eqref{eq:4} yields a negative D{\"o}ring mass. The instability arises with respect to a inhomogeneous perturbation localized at the DW plane (corrugation mode). \cite{MagyariThomas1985, Khodenkov2003}

\begin{figure}[!ht]
\includegraphics[width=8cm]{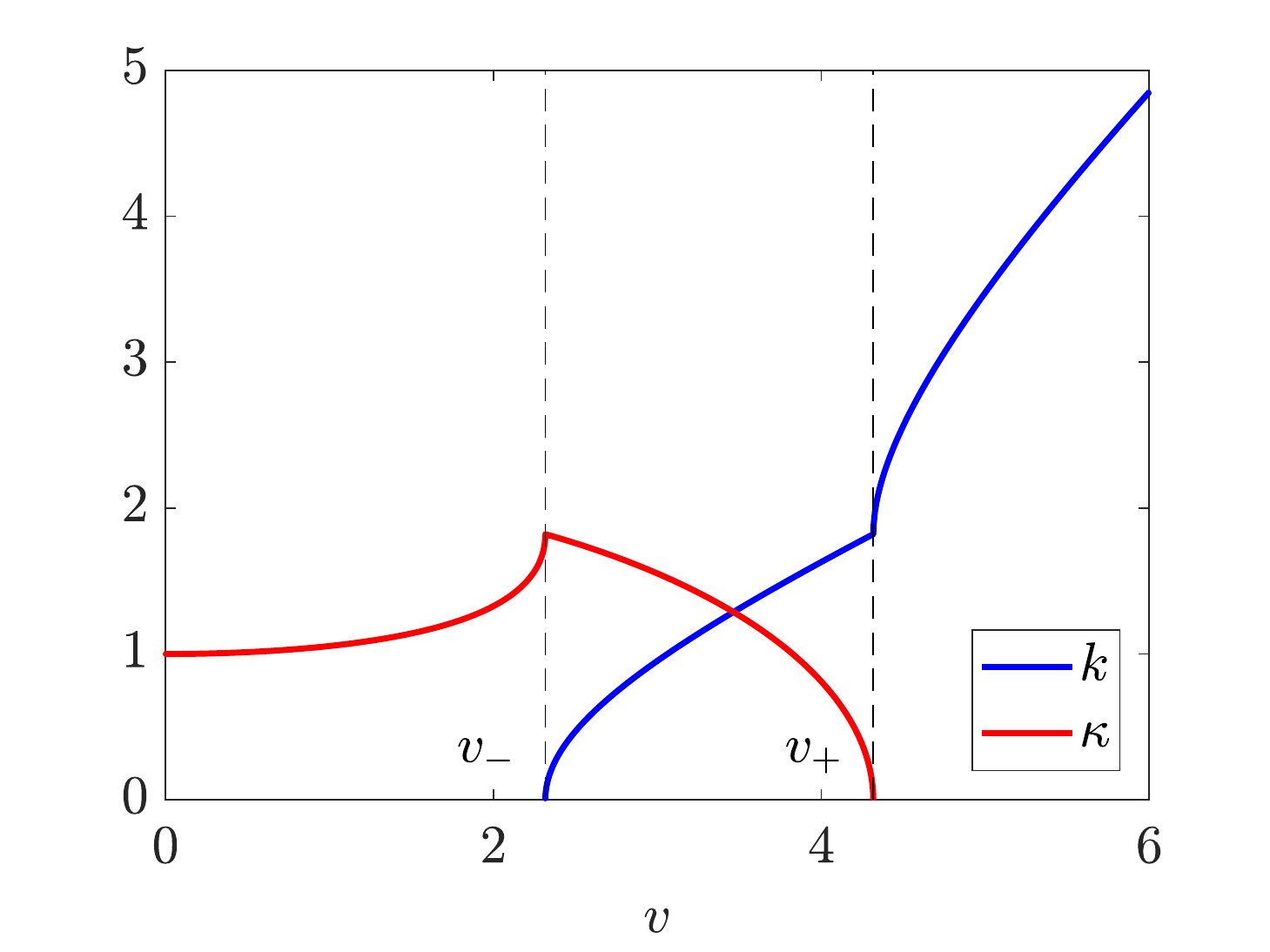}
\caption{Real $ k $ and imaginary $ \kappa $ wave vector components calculated by the generalization of the linear SW dispersion relation of biaxial FM as functions of the soliton velocity $v$ according to Eq. \eqref{eq:4} for $ \lambda = 10 $. The region $0<v <v_-$ corresponds to a moving DW.}
\label{im:2}
\end{figure}

As it has been pointed out, the critical velocities $ v_{\pm} $ separate regions that hold different moving  magnetization textures. Taking advantage of the parameterization of the DW dynamics in the region $ \left [0, v_- \right] $ through Eq. \eqref{eq:4}, it is possible to discuss how well this generalization adapts to the rest of the velocity regions. This has been done in Fig. \ref{im:2}. In the interval $ \left[ 0, v_- \right] $, being the wave vector $ K = \di \kappa $ and frequency $ \Omega = \di \kappa v $ purely imaginary, the magnetization waves are localized spatially forming a DW. The DW width contracts as $ v $ increases but its structure remains unchanged and the DW non-zero topological charge is conserved, at least until reaching the limiting velocity $ v_- $. In the velocity domain $ \left( v_-, v_+ \right) $, where both $ K $ and $ \Omega $ are complex, the real component of the wave vector $k$ appears above the velocity $ v_- $ and increases as $v$ increases, and the imaginary component ${\kappa}$ decreases to be zero at the critical velocity $ v_+ $.  According to Ref. [\citen{KosevichIvanovKovalev1990}], the magnetization profile within the region  $ \left( v_-, v_+ \right) $ can be described as a localized envelope of the non-topological soliton (the area of localization is $1/{\kappa}$) modulated by a periodic pattern with the wavelength about of $1/k$ (that is, some SW oscillations appear along the soliton envelope). At this point, we would like to emphasize that when the term "soliton" is used in the absence of specifications on its topological charge in this text, we will be referring to non-topological solitons, while, on the other hand, DW constitutes topological ones. Finally, in the last region $\left[ \left. v_+, \infty \right) \right.$, only non-linear SW would be expected, obtaining the logical analogue in our case within the linear SW approximation.

\section{Mapping to a sine-Gordon equation}\label{section:gordon}

The Lorentz-invariant SG equation, having the exact $N$-soliton solutions, \cite{Faddeev1974} is one of the simplest non-linear equations. The dynamics of FM cannot be described, in general, by Lorentz-invariant equations, unlike in the case of AFM dynamics, where it is naturally described by a relativistic SG equation within the non-linear $\sigma$-model if the Zeeman and Dzyaloshinskii-Moriya interactions are absent. \cite{IvanovKolezhuk1995} Is it possible to reduce the non-linear LL equations, given by Eqs. \eqref{eq:2}, to the simple non-linear SG equation? Is it really necessary to consider an extremely large easy-plane anisotropy to achieve this mapping? \cite{BraunLoss1996, wieser2010domain}
To answer these questions, we introduce a new variable $ \eta $ defined as $ \tan \theta / 2 = \mathrm{exp} \left(- \eta \right)$ (such substitution is often used in the theory of magnetic solitons). \cite{KosevichIvanovKovalev1990} This allows to rewrite Eqs. \eqref{eq:2} as
\begin{equation}
\begin{aligned}
v \eta_{\xi}= \lambda \cos \phi \sin \phi + \phi_{\xi \xi} -2 \phi_{\xi} \eta_{\xi} \tanh \eta , \\
-v \phi_{\xi}= \eta_{\xi \xi}+ \left[ 1+\lambda \cos^2 \phi +{\left( \phi_{\xi} \right)}^2-{\left( \eta_{\xi} \right)}^2 \right] \tanh \eta .
\end{aligned}
\label{eq:5}
\end{equation}

The standard approach to this problem is to consider steady motion Walker-type of solutions, assuming $ \phi_{\xi} = 0 $, $\phi \left( \xi \right) = \phi_0$. \cite{SchryerWalker1974} This simplification leads to the system of equations given by $v \eta_{\xi}=\lambda \cos \phi_0 \, \sin \phi_0$, 
and $1+\lambda \cos^2 \phi_0 ={\left( \eta_{\xi} \right)}^2$. A stable solution of the form $\eta \left( \xi \right)= \xi / \Delta \left( v \right)$ exists when the DW velocity $v$ does not exceed $v_-$. Interestingly, the steady DW motion with a constant $\phi_0$ corresponds to the conservation of the magnetization field momentum $P=2\phi_0$. \cite{KosevichIvanovKovalev1990}  

The linear dependence $\eta \left( \xi \right)$ leads to the equation $\theta_{\xi \xi}= \sin 2 \theta / \,  2 \Delta^2 $, the same as in the static case except for the velocity-dependent DW width, $\Delta \left( v \right) = 1/ \kappa \left( v \right)$. Although the exact solution of the LL equation exists for the Walker case $ \phi_{\xi} = 0 $, the equation for the angle $\theta$ can be formally written as a SG equation without any assumptions about the value of the parameter $\lambda$, giving rise to
\begin{equation}
\theta_{xx}-\frac{1}{v^2_-} \, \ddot{\theta}=\frac{1}{2 \Delta^2_e} \sin 2 \theta ,
\label{eq:6}
\end{equation}
where $\Delta_e=\Delta / \sqrt{1-{\left( v /v_- \right)}^2}$, playing $v_-$ the role of the maximum DW velocity. Although the Walker-type solution $\phi \left( \xi \right) = \phi_0$ is stable, it is not possible to reproduce this result through an effective relativistic Lagrangian because the kinetic part of the Lagrangian density, $\mathcal{L}_{\mathrm{kin}}= - \phi \, \dot{\theta} \sin \theta$,  does not contribute to the equation of motion due to the simplified form of the solution considered.  Therefore, the formal SG equation given by Eq. \eqref{eq:6} is nonphysical and another mapping should be found. 

On the other hand, the limit $ \lambda \gg 1$ can be studied. Such limit can be realized even in soft magnetic materials like permalloy (NiFe alloy) or YIG with an induced uniaxial magnetic anisotropy $K_z$, where $\lambda = 2 \pi M^2_s / K_z >>1$, $K_x = 0$. For instance, a ratio $ \lambda = 21$ was used by Schryer  {\it et al}. for YIG. \cite{SchryerWalker1974} We assume that within the limit $ \lambda \gg 1$ the magnetization component $m_x$, perpendicular to the ``easy''-plane, is small, $m_x\ll 1$, and develop a perturbation theory with respect to it. A new scalar field, $\psi \left( x,t \right)$, is introduced for convenience through the equations $m_y=\sqrt{1-m^2_x} \, \sin \psi$, and $m_z= \sqrt{1-m^2_x} \, \cos \psi$. There are different ways to define $m_x$. We choose the {\it ansatz} $m_x= \varepsilon \sin \psi$, being $\varepsilon=\pi/2-\phi$ as sketched in Fig. \ref{im:1}, and we assume that $\varepsilon \left( \lambda \right) \rightarrow 0$ at $\lambda \gg 1$. In general, $\varepsilon$ does not have to be small and may be a function of $x$ and $t$. The initial variables ($\theta, \phi$) are related to new ones ($\psi, \varepsilon$) by the expressions $\cos \theta =\sqrt{1-\varepsilon^2 \sin^2 \psi} \, \cos \psi$, and $\cos \phi = \varepsilon / \sqrt{1+\varepsilon^2 \cos^2 \psi}$. If $\varepsilon \rightarrow 0$, $\cos \phi = \varepsilon +\mathcal{O} \left( \varepsilon^3 \right)$, and $\sin \phi =1+ \mathcal{O} \left( \varepsilon^2 \right)$. Substituting these expressions in Eqs. \eqref{eq:5} and assuming that $\varepsilon \left( \xi \right) =\mathrm{const}$, we obtain $v \eta_{\xi}= \lambda \varepsilon$, $\eta_{\xi \xi}=0$, and $1+ \lambda \varepsilon^2= {\left( \lambda \varepsilon / v \right)}^2$. The last equation has the solution $\varepsilon \left( \lambda, v \right)=v/ \lambda \sqrt{1-v^2 / \lambda}$. At this point, a new critical velocity can be introduced, $c= \sqrt{\lambda}$. It is straightforward to show that $\displaystyle c= \lim_{\lambda \gg 1} v_- \left( \lambda \right)$. The variable $\psi \left( x, t \right)$ coincides with the polar angle, $\theta \left( x, t \right)$, in the limit $\varepsilon \ll 1 $, if the terms of order $\mathcal{O} \left( \varepsilon^2 \right)$ are neglected. In this context, accounting for the approximate solution $\eta \left( \xi \right) = \xi / \Delta^{\prime} \left( v \right) = \xi / \sqrt{1-v^2/c^2}$, the SG equation for the DW profile angle $\theta \left( x, t \right) = 2 \arctan \mathrm{exp} \left(- \eta \left( \xi \right) \right)$ can be deduced
\begin{equation}
\theta_{xx}- \frac{1}{c^2} \ddot{\theta}=\frac{1}{2} \sin 2 \theta,
\label{eq:7}
\end{equation}
which corresponds to the particular case of Eq. \eqref{eq:6} in which $\Delta=\Delta^{\prime}$ and $v_-=c$. The SG-like expression given by Eq. \eqref{eq:7} is valid not only in the case of a Bloch-type DW that propagates in the direction perpendicular to the easy-plane, as it has been considered in our case, but also for, for example, the case of a N\'eel-type DW that propagates parallel to the direction dictated by the easy-axis symmetry-breaking anisotropy imposed in the system. \cite{hill2018spin}

\begin{figure}[!ht]
\includegraphics[width=8cm]{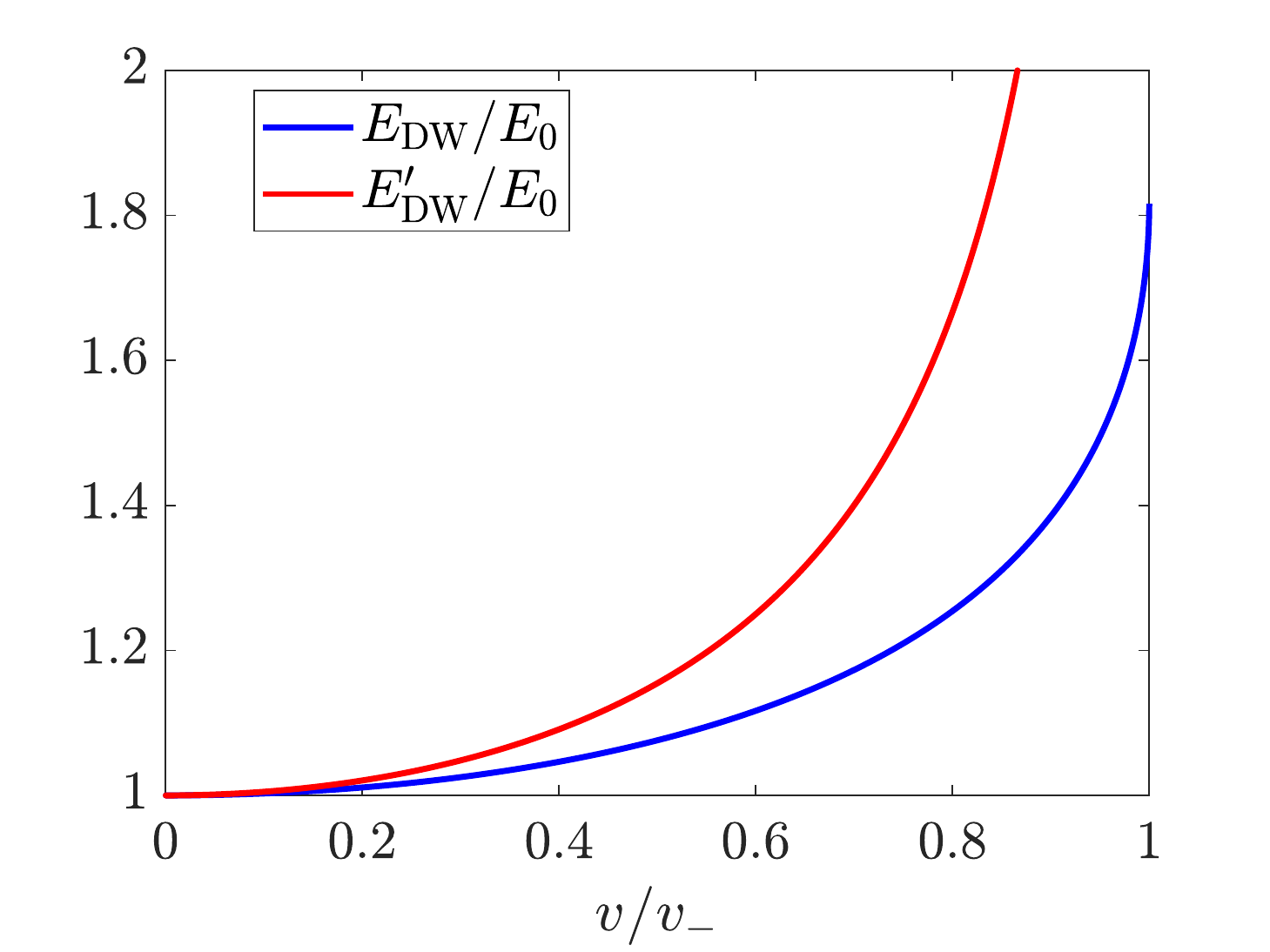}
\caption{Comparison between the moving DW energy obtained by using Eq. \eqref{eq:4}, $ E_{\mathrm{DW}} / E_0 $, for $ \lambda = 10 $, and a relativistic Lorentz-like energy, $E^{\prime}_{\mathrm{DW}} /E_0$, for which the maximum speed is given by the maximum SW phase velocity $ v_- $.}
\label{im:3}
\end{figure}

Therefore, in the limit $ \lambda \gg 1$, the DW energy and the DW width can be represented in a relativistic-like form
\begin{equation}
E^{\prime}_{\mathrm{DW}}\left( v \right)= \frac{E_0}{\sqrt{1-v^2/c^2}}, \, \Delta^{\prime} \left( v \right)= \sqrt{1-v^2/c^2}.
\label{eq:8}
\end{equation}
The DW energy found by Eq. \eqref{eq:4} smoothly increases as the velocity do so. However, this increasing behavior is far from being that of the normalized relativistic-like form, $E^{\prime}_{\mathrm{DW}}/E_0$, exposed in  Eqs. \eqref{eq:8}, which accentuates the difference of the discussed dynamic equations with a SG equation for any finite $\lambda$, as it can be seen in Fig. \ref{im:3}. The correct energy decomposition can be obtained from Eqs. \eqref{eq:8} only when the DW velocity is small ($v^2/c^2 \ll 1$), $E^{\prime}_{\mathrm{DW}}\left( v \right)=E_0+E_0v^2/2c^2$, leading to the DW D{\"o}ring mass $m_{\mathrm{DW}}=E_0 / c^2$, which coincides with the one defined above. However, in the limit $v \rightarrow c $,  the energy is singular, $E^{\prime}_{\mathrm{DW}}\left( v \right) \rightarrow \infty$, and the DW width matches the ultimate case of the Lorentz contraction, $\Delta^{\prime} \left( v \right) \rightarrow 0$. These results are nonphysical because the parameter $\varepsilon \left( \lambda, v \right)=v/ \lambda \sqrt{1-v^2 / \lambda}$ diverges at $v \rightarrow c$, and cannot be considered as a small parameter anymore. The exact solutions of Eqs. (\ref{eq:2},~\ref{eq:3}) predict the finite DW energy $E_{\mathrm{DW}} \left( v_- \right) = E_0 \, {\left( 1 + \lambda \right)}^{1/4}$ and finite DW width $\Delta \left( v_- \right) ={\left( 1+ \lambda \right)}^{-1/4}$ at $v \rightarrow v_-$. Thus, although one can write a SG-like expression by means of Eq. \eqref{eq:7} and the relativistic-like Eqs. \eqref{eq:8}, they are strictly valid only in the limit of small DW velocity $v^2/c^2 \ll 1$. However, Eqs. \eqref{eq:8} are very good approximation at $ \lambda \gg 1$ if the DW velocity is far enough from the maximal velocity, $c$. The subsequent conclusion is that the approximate solution obtained for the case in which $\varepsilon=\textrm{const}$, given by Eqs. \eqref{eq:8}, is asymptotically exact when $\lambda \gg 1$, provided that $v^2/c^2 \ll 1$, being far from the singularity that this solution presents, becoming virtually indistinguishable from the exact solution in biaxial FM obtained through Eq. \eqref{eq:4}. 

There is another approximate solution of the system of Eqs. \eqref{eq:2} or \eqref{eq:5} in the limit $\lambda \gg 1$ assuming Walker-type of solutions which was proposed by Sklyanin, \cite{Sklyanin1979} $\phi \left( \xi \right) = \mathrm{const}$. The solution was found to have the form $m_x \left( \xi^{\prime} \right)= \varepsilon_s \sin \Psi \left( \xi^{\prime} \right)$, being $\varepsilon_s= V / R \, \Delta_s$, $\Psi \left( \xi^{\prime} \right)= 2 \arctan \mathrm{exp} \left( \xi^{\prime} / \Delta_s \right)$, and $\xi^{\prime}=x-Vt$, $V=\sqrt{R} \, v$, $\Delta_s = \sqrt{R / \gamma^{\prime}} \, \sqrt{1-{\left( V / V_0 \right)}^2}$, and $V_0= \sqrt{R} \,$, in the aforementioned limit $R^2 / \gamma^{\prime}= \lambda \rightarrow \infty$. The function $\Psi \left( x, t \right)$ satisfies a SG equation of the form 
\begin{equation}
\Psi_{xx}- \frac{1}{V_0^2} \, \ddot{\Psi}= \frac{\gamma^{\prime}}{2R}  \sin 2 \Psi,
\label{eq:9}
\end{equation}
which constitutes a particular case of Eq. \eqref{eq:6} at $\Delta_e = \sqrt{R / \gamma^{\prime}}$. The DW width $\Delta_s$ goes to infinity for any finite DW velocity $V < V_0$. The maximum DW velocity, $V_0 ={\left( \gamma^{\prime} \lambda \right)}^{1/4}$, is not correct (it should be equal to $v_-= \sqrt{\lambda}$). A redefinition of the value of the parameter $\gamma^{\prime}$ was proposed by Kivshar {\it et al}., being $\gamma^{\prime} =1$, \cite{KivsharMalomed1989} keeping in this way the same form of the SG-like expression given by Eq. \eqref{eq:9}.

The small parameter $\varepsilon_s$ is similar to the previously defined one, $\varepsilon = v / \lambda \, \Delta^{\prime} \left( v \right) $. Both of them are singular as the velocity $v$ approaches the critical value $V_0 ={\lambda}^{1/4}$ or $v_-= \sqrt{\lambda}$. That is a peculiarity of all solutions of Eq. \eqref{eq:6} with $\phi(\xi)=\mathrm{const}$. However, it must be faced that the solution studied, proposed by Sklyanin, \cite{Sklyanin1979} is nonphysical, since it corresponds to the vanishing right-hand side in Eq. \eqref{eq:9} and infinitely wide DW at $V<V_0$. Moreover, the Sklyanin's  solution  $\Psi (x, t)$ does not satisfy the LL equation, which in the limit $\lambda \rightarrow \infty$ is reduced to the SG-like expression given by Eq. \eqref{eq:7}. Therefore, Eq. \eqref{eq:9} is not correct. This is due to that the Sklyanin's assumption that $K_x$ is proportional to $R$ and that $K_z$ is proportional to $1/R$ and goes to $0$ at $R \rightarrow \infty$ is physically senseless. The correct limit corresponds to consider the ratio $K_x/K_z = \lambda \gg 1$  keeping a finite value of the anisotropy constant $K_z$. Otherwise, the DW separating two domains with opposite magnetizations directed along $z$-{\it th} axis disappears. A finite value of $K_z$ allows the proper normalization of the DW energy in the units of  $2K_z \Delta_0$  and DW width in the units of $\Delta_0$. 

Therefore, accounting the drawbacks of the approach with $\varepsilon \left( \xi \right) =\mathrm{const}$, to properly calculate the limit $\lambda \gg 1$, $m_x \ll 1$ at finite DW velocity $v < v_{-}$ and get a SG equation for the polar angle $\theta$, we need to assume that the azimuthal angle $\varepsilon \left( \xi \right) \ll 1$ is a function of the coordinate and time, and solve Eqs. \eqref{eq:5}. Following this approach the Walker-type solutions $\varepsilon \left( \xi \right) = \mathrm{const}$ considered above can be refined, and the high velocity singularity of the steady DW solution, $\varepsilon \left( \xi , v \right)$ at $v \rightarrow c$, disappears. However, any solution with variable angle $\varepsilon \left( \xi \right)$ is beyond the current theory of 1D topological magnetic solitons and will be considered elsewhere. It is necessary to emphasize, at this point, that all the works we have found within the framework of the mapping of the LL equations to a SG-like expression operate within the perturbation approach $\varepsilon \left( x, t \right) \ll 1$, and fail to incorporate the traveling wave \textit{ansatz} into their formalism. \cite{mikeska1981solitons, Enz1964, DoddMorrisEilbeckEtAl1982, hill2018spin} This is because, in order to obtain the aforementioned mapping, they consider that the spatial derivative of an angular variable equivalent to our $ \varepsilon $ is zero, while its time derivative is finite. Therefore, the approach developed above assuming $\varepsilon \left( \xi \right) =\mathrm{const}$, $\varepsilon \ll 1$, although it is not completely correct at large velocities $v \rightarrow c$, constitutes one step further than the approaches employed so far.

We note that the SG equation given by Eq. \eqref{eq:6} requires a kinetic Lagrangian density term of the form $\mathcal{L}_{\mathrm{kin}} \propto \dot{\theta}^2$, which is similar to the one for AFM within the non-linear $\sigma$-model giving rise to relativistic DW dynamics in this kind of systems. \cite{IvanovKolezhuk1995} To understand the appearance of such term, the kinetic Lagrangian density term for FM, $\mathcal{L}_{\mathrm{kin}}= -\dot{\phi} \cos \theta$, can be rewritten in the equivalent form $\mathcal{L}_{\mathrm{kin}}=- \phi \, \dot{\theta} \sin \theta$. Evaluating Eqs. \eqref{eq:2} at $\phi \left( \xi \right)= \pi /2 - \varepsilon \left( \xi \right)$, we find that the term $ \sin \theta $ is proportional to the time derivative $ \dot{\theta} $, namely $ \sin \theta = \dot{\theta} / \lambda \varepsilon $, which means that the kinetic Lagrangian density term can be expressed in a mass-like form, $\mathcal{L}_{\mathrm{kin}}= \dot{\theta}^2/ \lambda$. The effective Lagrangian density, $\mathcal{L}=e \left( \theta, \phi \right) - \mathcal{L}_{\mathrm{kin}}$, within the limit $m_x \ll 1$ is given by the expression
\begin{equation}
\mathcal{L}= {\left( \theta_x \right)}^2 +\sin^2 \theta - \frac{1}{c^2} \, \dot{\theta}^2,
\label{eq:10}
\end{equation}
which is compatible with the SG equation.

On the other hand, for an uniaxial AFM within the non-linear $\sigma$-model, the SW dispersion relation is $\omega^2=\omega^2_0+c^2k^2$, where $\omega_0$ is a frequency gap due to the uniaxial magnetic anisotropy. Employing the formalism of the complex wave vectors, $K=k+ \di \kappa$, it can be proved that $v_-=v_+=c$, and that the dependence $\kappa \left( v \right)$, in units of $\kappa_0=\omega_0 /c$, is expressed as $\kappa \left( v \right) = 1/ \sqrt{1-v^2/c^2}$. This immediately leads to relativistic-like expressions for the DW energy and DW width of AFM, $\Delta \left( v \right) = 1/ \kappa \left( v \right)$, similarly as to Eqs. \eqref{eq:8}, being
\begin{equation}
E_{\mathrm{DW}}\left( v \right)= \frac{E_0}{\sqrt{1-v^2/c^2}}, \, \Delta \left( v \right)=\Delta_0 \sqrt{1-v^2/c^2}.
\label{eq:11}
\end{equation}

However, in comparison to the approximate Eqs. \eqref{eq:8} for biaxial FM, Eqs. \eqref{eq:11} are exact within the non-linear $\sigma$-model of an uniaxial AFM. \cite{Fradkin2013} The SW velocity $c$ in AFM, limiting the DW velocity, is essentially higher than the one in FM due to the exchange enhancement. 

We consider that the mapping of the LL equation into a SG-like equation for biaxial FM results in more deep understanding of the DW dynamics. In particular, the DW velocity increase along with the Lorentz contraction of DW width (as a result of the SG equation of DW motion) can lead to considerable spin Peltier effect not only in specific AFM-like $\textrm{Mn}_2 \textrm{Au}$, \cite{OtxoaAtxitiaRoyEtAl2019}  but also in traditional FM metals with biaxial magnetic anisotropy. 

\section{Relativistic-like signatures in atomistic simulations}\label{section:simulations}

As it was previously introduced in Sec. \ref{section:limit}, there is a dynamical contraction of the DW width as it travels through a biaxial FM. Also, the higher the value of the magnetic anisotropy constants ratio $\lambda $, the greater this process will be. With this in mind, and in the spirit of the search for signatures of relativistic-like behaviors of the DW energy and DW width as was obtained in Eqs. \eqref{eq:8} for FM, we investigate this process numerically.  To do this, we studied, through atomistic spin dynamic simulations (fifth order Runge-Kutta method to solve numerically the Landau-Lifshitz-Gilbert equation site by site), how the velocity of the magnetic texture, $ v $, and the DW width, $ \Delta $, behaves, as the applied magnetic field, $ H $, directed along the anisotropy easy-axis, is increased. We exploited atomistic spin dynamics simulations because the DW width for high velocities is expected to be about of $1$ nm and the continuous approach fails. With this goal in mind, the simulated system will be given by a 1D FM spin chain consisting of $ 60000$ atomic sites. We use the typical magnetic parameters for the FM layers that make up the layered AFM  $\textrm{Mn}_2 \textrm{Au} $, \cite{BarthemColinMayaffreEtAl2013} being the exchange integral $ I= 1.588 \cdot 10^{-21}$ J, the atomic moment $\mu = 4 \, \mu_B$, and the lattice period $a_0=0.3328$ nm. In addition, the hard-axis anisotropy constant is given by $K_{x} \, a_0^3 = I$, the easy-axis anisotropy constant possess the value $K_z \, a_0^3=1.302 \cdot 10^{-24}$ J, the gyromagnetic ratio is equal to $\gamma=2.21 \cdot 10^5$ m/(A$\cdot$s), and the Gilbert damping constant is expressed by $\alpha =0.001$. These particular values of the anisotropy constants were chosen to secure the limit $ m_x \ll 1 $, $\varepsilon \left( \xi \right) \ll 1$.

To find moving DW solutions for the case of non-zero magnetic field and non-zero damping we include the corresponding terms to Eqs. \eqref{eq:2}. We consider only the Walker-type of solution $\phi \left( \xi \right) = \mathrm{const}$, which gives rise to
\begin{equation}
\begin{aligned}
 \dot{\theta}= \lambda \cos  \phi  \sin  \phi  \sin  \theta ,
 \\
 \alpha \, \dot{\theta} - h \sin \theta=\left( 1+\lambda \cos^2  \phi  \right)  \cos  \theta  \sin  \theta -\theta_{xx},
\label{eq:12}
\end{aligned}
\end{equation}
where $h=H/H_a$ is the reduced external magnetic field.

We want to keep the kink solution of Eqs. (\ref{eq:2},~\ref{eq:5}) unchanged and search for a specific Walker-type of solution assuming that the damping- and field-terms cancel each other, i.e.,  $\alpha \, \dot{\theta} = h \, \sin \theta$. Accounting that for the kink solution, $v / \Delta (v)=\lambda \cos \phi_0 \sin \phi_0$, the equation connecting the DW velocity and magnetic field for any value of $\lambda$ is given by
\begin{equation}
\frac{ v } {\Delta (v)} = \frac {h} {\alpha},
\label{eq:13}
\end{equation}
where the velocity-dependent DW width $\Delta(v) = 1/ \kappa(v)$ can be determined from Eq. \eqref{eq:4}. 

Now, to evaluate how well the data thrown by the simulations fits into a relativistic-like behavior which is a result of the SG equation of motion, it is necessary to take into account the DW contraction $\Delta(v) $ as the velocity increases. The DW width dependence on the DW velocity for the limiting case $ \lambda \gg 1$ can be approximately described by Eqs. \eqref{eq:8} as $ \Delta(v) = \Delta_0 \sqrt{1-v^2 / c^2} $, where $ \Delta_0 $ is the DW width at rest, which in the simulations is $ \Delta_0 = 11.62 $ nm, and $c$ is the maximal DW velocity, which was simulated to be $c=4.981$ km/s. Recalculating the atomistic parameters to the micromagnetic ones, we get $M_s=1006$ kA/m, $A=4.77$ pJ/m,  $K_x=43.08$ $\textrm{MJ/m}^3$, $K_z= 0.0353$ $\textrm{MJ/m}^3$, and extremely large $\lambda=1220$. The maximal steady DW velocity then is $v_-=4.869$ km/s. This value is very close to the simulated maximal value of the DW velocity, $c$.  

The equation for the DW velocity $v(h)$ can be easily solved for the ultimate case $ \lambda \gg 1$. The expression of velocity as a function of the magnetic field is explicitly given by
\begin{equation}
v(h)= \frac{h/ \alpha}{\sqrt{1+(h/ \alpha c)^2}}.
\label{eq:14}
\end{equation}

\begin{figure}[!ht]
\includegraphics[width=8cm]{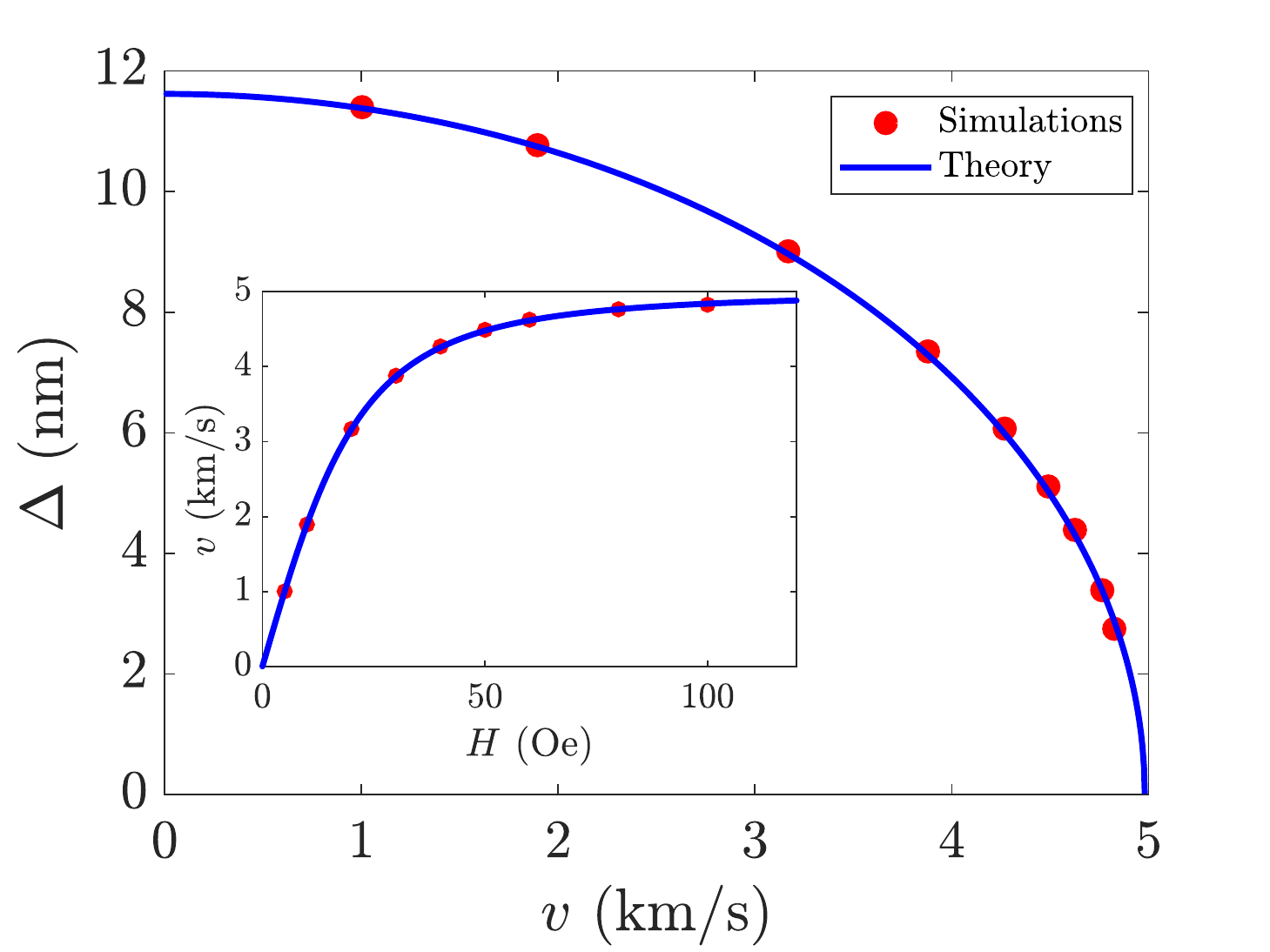}
\caption{Atomistic spin dynamics simulations for a FM simple cubic lattice using the parameters listed in the text, from which a relativistic-like behavior can be extracted for the DW width $ \Delta $ (which contracts as the velocity increases) and for the DW velocity $ v $ (which saturates as the external magnetic field increases).}
\label{im:4}
\end{figure}

As it can be seen in Fig. \ref{im:4}, there is a very good match between the simulations and what is predicted by the theory in the limit $ \lambda \gg 1$. We note that for small fields (velocities) the expression for the DW velocity coincides with the standard Walker expression derived in the limit $\lambda =0$. \cite{SchryerWalker1974} On the other side, the expression for $v(h)$ has the same form as the expression for DW velocity in weak FM-like ${\textrm{YFeO}}_3$. \cite{bar2006dynamics} Only the DW mobility $v/H$ is different, because in weak FM it is determined by the exchange and Dzyaloshinskii-Moriya interactions.   
Therefore, this endorses the idea that, in extreme case of biaxial FM, it is possible to obtain traces of the behavior that characterizes weak FM (AFM), since the DW velocity saturates as the applied magnetic field increases and the DW width contracts drastically as the velocity of the magnetic texture increases. The DW steady motion velocity $v(h)$ is not an arbitrary parameter as it was assumed in previous sections.  The steady-state DW motion of this new type of Walker solutions is possible only for a definite value of the velocity which is determined by the given magnetic field and damping parameter according to the equation for $v(h)$. Good agreement between the simulated dependencies $ \Delta(v) $, $v(h)$ and the ones calculated within the Walker approximation $\phi(\xi)=\textrm{const}$ means that the effect of the variable $\varepsilon \left( \xi \right) \ll 1$ on the DW dynamics is small for the very large $\lambda \gg 1$ in a wide DW velocity region up to $v_-$. 

In fact, small values of the DW widths at high velocities give rise to a fundamental question. In micromagnetic simulations, which are usually used to evaluate the dynamics of magnetic textures in FM, a certain cell size must be chosen at the beginning of the numerical process. However, if the  DW width contracts, it could be the case that, at a certain moment, the cell size chosen is insufficient to capture all the physics present in the problem, which could lead to artificial results that do not correspond to reality. Therefore, it is important to draw attention to this fact, since it could have an enormous impact in some cases. A perhaps more precise way to work in the context of micromagnetic simulations would be to apply a correction to  the cell size according to the effect that a Lorentz-like factor could have on the DW width in the medium.

\section{Conclusions}\label{section:conclusions}

We addressed  the problem of the reduction of the LL equation of motion to the SG equation in a bulk biaxial FM. The investigation on how to increase the DW velocity of anisotropic FM and whether such velocity increase is related to the SG equation of magnetization motion is of considerable importance.  In the framework of steady-state Walker-type of solutions, it is formally possible to obtain the aforementioned Lorentz-invariant SG equation for the general case of arbitrary $\lambda$ for any DW velocity $ v < v_- $. However, the mapping has physical sense only in the limit of extremely large ratio of easy-plane and easy-axis magnetic anisotropy constants $ \lambda \gg 1 $ for DW velocities $ v \ll v_- $. The singularities were found for the later case in which the magnetization angle  $ \varepsilon = \pi / 2 - \phi $ was operated in the limit $ \varepsilon \ll 1 $. We believe that accounting for the spatial and time dependence of the variable $ \varepsilon $ is sufficient to avoid the singularities in the limit $\lambda \gg 1$ and get the Lorentz-invariant Lagrangian along with SG equation for the magnetization polar angle $\theta (\xi)$. The possibility of mapping the LL equation into the SG equation does not imply an increase of the maximum DW velocity over the Walker-type solution velocity limit $v_-$. The case of the variable azimuthal magnetization angle $\phi (\xi)$ or even the essentially simpler case of the variable $ \varepsilon (\xi) \ll 1 $ are beyond of the theory of 1D magnetic solitons and will be considered elsewhere. The 1D DW dynamics considered above can be used not only for the description of bulk anisotropic FM and 1D spin chains, but also for 2D magnetic systems such as nanowires and nanostripes with small cross-section, if the conditions of applicability of 1D DW model discussed in Refs. [\citen{ThiavilleRohartJueEtAl2012, ThiavilleNakatani2006}] are satisfied. Moreover, as it has been proven through atomistic spin dynamics simulations, it is in fact possible to replicate through the precepts of special relativity the behavior of the dynamics of a DW in a biaxial FM for the case $ \lambda \gg 1 $. Such a remarkable result supports what is analytically predicted throughout this text, and shows that, if there is a FM material for which $ \lambda $ was large enough, a system could be experimentally implemented in which WB-induced instabilities disappear without using complicated geometries. Furthermore, as mentioned, the selection of cell size in micromagnetic simulations requires a deeper reconsideration. In the high-speed regime, due to the DW width contraction, the chosen cell size may be insufficient to consider that the continuum approximation continues to be satisfied, which could cause the exchange interaction between the spins that make up the DW to be wrong, due to that the angle between spins would actually be smaller than the micromagnetic simulations would predict, leading to a more abrupt unreal transition through the DW. Because of this, it would be necessary to introduce a dynamic exchange length that took into account relativistic effects at high speeds. As a result, atomistic spin dynamics simulations would generally yield a more precise result even in the case of FM. 

\section{Acknowledgements}\label{section:acknowledgements}

K. G. acknowledges support by IKERBASQUE (the Basque Foundation for Science) and Spanish MINECO project FIS2016-78591-C3-3-R. The work of R. O. and K. G. was partially supported by the STSM Grants from the COST Action CA17123 ``Ultrafast opto-magneto-electronics for non-dissipative information technology".

\bibliography{SineGordonBibtex}

\begin{thebibliography}{32}
\expandafter\ifx\csname natexlab\endcsname\relax\def\natexlab#1{#1}\fi
\expandafter\ifx\csname bibnamefont\endcsname\relax
  \def\bibnamefont#1{#1}\fi
\expandafter\ifx\csname bibfnamefont\endcsname\relax
  \def\bibfnamefont#1{#1}\fi
\expandafter\ifx\csname citenamefont\endcsname\relax
  \def\citenamefont#1{#1}\fi
\expandafter\ifx\csname url\endcsname\relax
  \def\url#1{\texttt{#1}}\fi
\expandafter\ifx\csname urlprefix\endcsname\relax\def\urlprefix{URL }\fi
\providecommand{\bibinfo}[2]{#2}
\providecommand{\eprint}[2][]{\url{#2}}

\bibitem[{\citenamefont{Landau and Lifshitz}(1935)}]{LandauLifshitz1935}
\bibinfo{author}{\bibfnamefont{L.}~\bibnamefont{Landau}} \bibnamefont{and}
  \bibinfo{author}{\bibfnamefont{E.}~\bibnamefont{Lifshitz}},
  \bibinfo{journal}{Phys. Z. Sowjetunion} \textbf{\bibinfo{volume}{8}},
  \bibinfo{pages}{153} (\bibinfo{year}{1935}).

\bibitem[{\citenamefont{Gomonay et~al.}(2018)\citenamefont{Gomonay, Baltz,
  Brataas, and Tserkovnyak}}]{GomonayBaltzBrataasEtAl2018}
\bibinfo{author}{\bibfnamefont{O.}~\bibnamefont{Gomonay}},
  \bibinfo{author}{\bibfnamefont{V.}~\bibnamefont{Baltz}},
  \bibinfo{author}{\bibfnamefont{A.}~\bibnamefont{Brataas}}, \bibnamefont{and}
  \bibinfo{author}{\bibfnamefont{Y.}~\bibnamefont{Tserkovnyak}},
  \bibinfo{journal}{Nature Phys.} \textbf{\bibinfo{volume}{14}},
  \bibinfo{pages}{213} (\bibinfo{year}{2018}).

\bibitem[{\citenamefont{Kosevich et~al.}(1990)\citenamefont{Kosevich, Ivanov,
  and Kovalev}}]{KosevichIvanovKovalev1990}
\bibinfo{author}{\bibfnamefont{A.~M.} \bibnamefont{Kosevich}},
  \bibinfo{author}{\bibfnamefont{B.}~\bibnamefont{Ivanov}}, \bibnamefont{and}
  \bibinfo{author}{\bibfnamefont{A.}~\bibnamefont{Kovalev}},
  \bibinfo{journal}{Phys. Reports} \textbf{\bibinfo{volume}{194}},
  \bibinfo{pages}{117} (\bibinfo{year}{1990}).

\bibitem[{\citenamefont{Schryer and Walker}(1974)}]{SchryerWalker1974}
\bibinfo{author}{\bibfnamefont{N.~L.} \bibnamefont{Schryer}} \bibnamefont{and}
  \bibinfo{author}{\bibfnamefont{L.~R.} \bibnamefont{Walker}},
  \bibinfo{journal}{J. Appl. Phys.} \textbf{\bibinfo{volume}{45}},
  \bibinfo{pages}{5406} (\bibinfo{year}{1974}).

\bibitem[{\citenamefont{Fradkin}(2013)}]{Fradkin2013}
\bibinfo{author}{\bibfnamefont{E.}~\bibnamefont{Fradkin}},
  \emph{\bibinfo{title}{Field Theories of Condensed Matter Physics}}
  (\bibinfo{publisher}{Cambridge University Press}, \bibinfo{year}{2013}).

\bibitem[{\citenamefont{Shiino et~al.}(2016)\citenamefont{Shiino, Oh, Haney,
  Lee, Go, Park, and Lee}}]{shiino2016antiferromagnetic}
\bibinfo{author}{\bibfnamefont{T.}~\bibnamefont{Shiino}},
  \bibinfo{author}{\bibfnamefont{S.-H.} \bibnamefont{Oh}},
  \bibinfo{author}{\bibfnamefont{P.~M.} \bibnamefont{Haney}},
  \bibinfo{author}{\bibfnamefont{S.-W.} \bibnamefont{Lee}},
  \bibinfo{author}{\bibfnamefont{G.}~\bibnamefont{Go}},
  \bibinfo{author}{\bibfnamefont{B.-G.} \bibnamefont{Park}}, \bibnamefont{and}
  \bibinfo{author}{\bibfnamefont{K.-J.} \bibnamefont{Lee}},
  \bibinfo{journal}{Phys. Rev. Lett.} \textbf{\bibinfo{volume}{117}},
  \bibinfo{pages}{087203} (\bibinfo{year}{2016}).

\bibitem[{\citenamefont{Gomonay et~al.}(2016)\citenamefont{Gomonay, Jungwirth,
  and Sinova}}]{gomonay2016high}
\bibinfo{author}{\bibfnamefont{O.}~\bibnamefont{Gomonay}},
  \bibinfo{author}{\bibfnamefont{T.}~\bibnamefont{Jungwirth}},
  \bibnamefont{and} \bibinfo{author}{\bibfnamefont{J.}~\bibnamefont{Sinova}},
  \bibinfo{journal}{Phys. Rev. Lett.} \textbf{\bibinfo{volume}{117}},
  \bibinfo{pages}{017202} (\bibinfo{year}{2016}).

\bibitem[{\citenamefont{Wieser et~al.}(2010)\citenamefont{Wieser, Vedmedenko,
  and Wiesendanger}}]{wieser2010domain}
\bibinfo{author}{\bibfnamefont{R.}~\bibnamefont{Wieser}},
  \bibinfo{author}{\bibfnamefont{E.}~\bibnamefont{Vedmedenko}},
  \bibnamefont{and}
  \bibinfo{author}{\bibfnamefont{R.}~\bibnamefont{Wiesendanger}},
  \bibinfo{journal}{Phys. Rev. B} \textbf{\bibinfo{volume}{81}},
  \bibinfo{pages}{024405} (\bibinfo{year}{2010}).

\bibitem[{\citenamefont{Yan et~al.}(2010)\citenamefont{Yan, K{\'a}kay, Gliga,
  and Hertel}}]{YanHertelPRL2010}
\bibinfo{author}{\bibfnamefont{M.}~\bibnamefont{Yan}},
  \bibinfo{author}{\bibfnamefont{A.}~\bibnamefont{K{\'a}kay}},
  \bibinfo{author}{\bibfnamefont{S.}~\bibnamefont{Gliga}}, \bibnamefont{and}
  \bibinfo{author}{\bibfnamefont{R.}~\bibnamefont{Hertel}},
  \bibinfo{journal}{Phys. Rev. Lett.} \textbf{\bibinfo{volume}{104}},
  \bibinfo{pages}{057201} (\bibinfo{year}{2010}).

\bibitem[{\citenamefont{Yan et~al.}(2011)\citenamefont{Yan, Andreas, K{\'a}kay,
  Garc{\'\i}a-S{\'a}nchez, and Hertel}}]{YanAndreasKakayEtAl2011}
\bibinfo{author}{\bibfnamefont{M.}~\bibnamefont{Yan}},
  \bibinfo{author}{\bibfnamefont{C.}~\bibnamefont{Andreas}},
  \bibinfo{author}{\bibfnamefont{A.}~\bibnamefont{K{\'a}kay}},
  \bibinfo{author}{\bibfnamefont{F.}~\bibnamefont{Garc{\'\i}a-S{\'a}nchez}},
  \bibnamefont{and} \bibinfo{author}{\bibfnamefont{R.}~\bibnamefont{Hertel}},
  \bibinfo{journal}{Appl. Phys. Lett.} \textbf{\bibinfo{volume}{99}},
  \bibinfo{pages}{122505} (\bibinfo{year}{2011}).

\bibitem[{\citenamefont{Thiaville et~al.}(2012)\citenamefont{Thiaville, Rohart,
  Ju{\'e}, Cros, and Fert}}]{ThiavilleRohartJueEtAl2012}
\bibinfo{author}{\bibfnamefont{A.}~\bibnamefont{Thiaville}},
  \bibinfo{author}{\bibfnamefont{S.}~\bibnamefont{Rohart}},
  \bibinfo{author}{\bibfnamefont{{\'E}.}~\bibnamefont{Ju{\'e}}},
  \bibinfo{author}{\bibfnamefont{V.}~\bibnamefont{Cros}}, \bibnamefont{and}
  \bibinfo{author}{\bibfnamefont{A.}~\bibnamefont{Fert}},
  \bibinfo{journal}{Europhys. Lett.} \textbf{\bibinfo{volume}{100}},
  \bibinfo{pages}{57002} (\bibinfo{year}{2012}).

\bibitem[{\citenamefont{Mikeska}(1981)}]{mikeska1981solitons}
\bibinfo{author}{\bibfnamefont{H.}~\bibnamefont{Mikeska}}, \bibinfo{journal}{J.
  Appl. Phys.} \textbf{\bibinfo{volume}{52}}, \bibinfo{pages}{1950}
  (\bibinfo{year}{1981}).

\bibitem[{\citenamefont{Schl{\"o}mann}(1971)}]{Schloemann1971}
\bibinfo{author}{\bibfnamefont{E.}~\bibnamefont{Schl{\"o}mann}},
  \bibinfo{journal}{Appl. Phys. Lett.} \textbf{\bibinfo{volume}{19}},
  \bibinfo{pages}{274} (\bibinfo{year}{1971}).

\bibitem[{\citenamefont{Enz}(1964)}]{Enz1964}
\bibinfo{author}{\bibfnamefont{U.}~\bibnamefont{Enz}}, \bibinfo{journal}{Helv.
  Phys. Acta} \textbf{\bibinfo{volume}{37}}, \bibinfo{pages}{245}
  (\bibinfo{year}{1964}).

\bibitem[{\citenamefont{Dodd et~al.}(1982)\citenamefont{Dodd, Morris, Eilbeck,
  and Gibbon}}]{DoddMorrisEilbeckEtAl1982}
\bibinfo{author}{\bibfnamefont{R.~K.} \bibnamefont{Dodd}},
  \bibinfo{author}{\bibfnamefont{H.~C.} \bibnamefont{Morris}},
  \bibinfo{author}{\bibfnamefont{J.}~\bibnamefont{Eilbeck}}, \bibnamefont{and}
  \bibinfo{author}{\bibfnamefont{J.}~\bibnamefont{Gibbon}},
  \bibinfo{journal}{{\it Soliton and Nonlinear Wave Equations}, Academic Press,
  London and New York, 640 p., Chapt. 7}  (\bibinfo{year}{1982}).

\bibitem[{\citenamefont{D{\"o}ring}(1948)}]{Doering1948}
\bibinfo{author}{\bibfnamefont{W.}~\bibnamefont{D{\"o}ring}},
  \bibinfo{journal}{Zeitschrift f{\"u}r Naturforschung A}
  \textbf{\bibinfo{volume}{3}}, \bibinfo{pages}{373} (\bibinfo{year}{1948}).

\bibitem[{\citenamefont{Eleonskii et~al.}(1978)\citenamefont{Eleonskii, Kirova,
  and Kulagin}}]{EleonskiKirovaKulagin1978}
\bibinfo{author}{\bibfnamefont{V.}~\bibnamefont{Eleonskii}},
  \bibinfo{author}{\bibfnamefont{N.}~\bibnamefont{Kirova}}, \bibnamefont{and}
  \bibinfo{author}{\bibfnamefont{N.}~\bibnamefont{Kulagin}},
  \bibinfo{journal}{Soviet Journal of Experimental and Theoretical Physics}
  \textbf{\bibinfo{volume}{74}}, \bibinfo{pages}{1814} (\bibinfo{year}{1978}).

\bibitem[{\citenamefont{Sobolev et~al.}(1995)\citenamefont{Sobolev, Huang, and
  Chen}}]{SobolevHuangChen1995}
\bibinfo{author}{\bibfnamefont{V.}~\bibnamefont{Sobolev}},
  \bibinfo{author}{\bibfnamefont{H.}~\bibnamefont{Huang}}, \bibnamefont{and}
  \bibinfo{author}{\bibfnamefont{S.}~\bibnamefont{Chen}}, \bibinfo{journal}{J.
  Magn. Magn. Mat.} \textbf{\bibinfo{volume}{147}}, \bibinfo{pages}{284}
  (\bibinfo{year}{1995}).

\bibitem[{\citenamefont{Magyari and Thomas}(1985)}]{MagyariThomas1985}
\bibinfo{author}{\bibfnamefont{E.}~\bibnamefont{Magyari}} \bibnamefont{and}
  \bibinfo{author}{\bibfnamefont{H.}~\bibnamefont{Thomas}},
  \bibinfo{journal}{Zeitschrift f{\"u}r Physik B Cond. Matter}
  \textbf{\bibinfo{volume}{59}}, \bibinfo{pages}{167} (\bibinfo{year}{1985}).

\bibitem[{\citenamefont{Eleonskii et~al.}(1976)\citenamefont{Eleonskii, Kirova,
  and Kulagin}}]{EleonskiiKirovaKulagin1976}
\bibinfo{author}{\bibfnamefont{V.}~\bibnamefont{Eleonskii}},
  \bibinfo{author}{\bibfnamefont{N.}~\bibnamefont{Kirova}}, \bibnamefont{and}
  \bibinfo{author}{\bibfnamefont{N.}~\bibnamefont{Kulagin}},
  \bibinfo{journal}{Soviet Journal of Experimental and Theoretical Physics}
  \textbf{\bibinfo{volume}{44}}, \bibinfo{pages}{1239} (\bibinfo{year}{1976}).

\bibitem[{\citenamefont{Hubert and Sch{\"a}fer}(2008)}]{HubertSchaefer2008}
\bibinfo{author}{\bibfnamefont{A.}~\bibnamefont{Hubert}} \bibnamefont{and}
  \bibinfo{author}{\bibfnamefont{R.}~\bibnamefont{Sch{\"a}fer}},
  \emph{\bibinfo{title}{Magnetic Domains: the Analysis of Magnetic
  Microstructures}} (\bibinfo{publisher}{Springer Science \& Business Media},
  \bibinfo{year}{2008}).

\bibitem[{\citenamefont{Khodenkov}(2003)}]{Khodenkov2003}
\bibinfo{author}{\bibfnamefont{G.}~\bibnamefont{Khodenkov}},
  \bibinfo{journal}{Techn. Phys. Lett.} \textbf{\bibinfo{volume}{29}},
  \bibinfo{pages}{907} (\bibinfo{year}{2003}).

\bibitem[{\citenamefont{Faddeev et~al.}(1974)\citenamefont{Faddeev, Takhtajan,
  and Zakharov}}]{Faddeev1974}
\bibinfo{author}{\bibfnamefont{L.}~\bibnamefont{Faddeev}},
  \bibinfo{author}{\bibfnamefont{L.}~\bibnamefont{Takhtajan}},
  \bibnamefont{and} \bibinfo{author}{\bibfnamefont{V.}~\bibnamefont{Zakharov}},
  in \emph{\bibinfo{booktitle}{Sov. Phys. Dokl.}} (\bibinfo{year}{1974}),
  vol.~\bibinfo{volume}{19}, pp. \bibinfo{pages}{824--826}.

\bibitem[{\citenamefont{Ivanov and Kolezhuk}(1995)}]{IvanovKolezhuk1995}
\bibinfo{author}{\bibfnamefont{B.}~\bibnamefont{Ivanov}} \bibnamefont{and}
  \bibinfo{author}{\bibfnamefont{A.}~\bibnamefont{Kolezhuk}},
  \bibinfo{journal}{Low Temp. Phys.} \textbf{\bibinfo{volume}{21}},
  \bibinfo{pages}{355} (\bibinfo{year}{1995}).

\bibitem[{\citenamefont{Braun and Loss}(1996)}]{BraunLoss1996}
\bibinfo{author}{\bibfnamefont{H.-B.} \bibnamefont{Braun}} \bibnamefont{and}
  \bibinfo{author}{\bibfnamefont{D.}~\bibnamefont{Loss}},
  \bibinfo{journal}{Phys. Rev. B} \textbf{\bibinfo{volume}{53}},
  \bibinfo{pages}{3237} (\bibinfo{year}{1996}).

\bibitem[{\citenamefont{Hill et~al.}(2018)\citenamefont{Hill, Kim, and
  Tserkovnyak}}]{hill2018spin}
\bibinfo{author}{\bibfnamefont{D.}~\bibnamefont{Hill}},
  \bibinfo{author}{\bibfnamefont{S.~K.} \bibnamefont{Kim}}, \bibnamefont{and}
  \bibinfo{author}{\bibfnamefont{Y.}~\bibnamefont{Tserkovnyak}},
  \bibinfo{journal}{Phys. Rev. Lett.} \textbf{\bibinfo{volume}{121}},
  \bibinfo{pages}{037202} (\bibinfo{year}{2018}).

\bibitem[{\citenamefont{Sklyanin}(1979)}]{Sklyanin1979}
\bibinfo{author}{\bibfnamefont{E.~K.} \bibnamefont{Sklyanin}},
  \bibinfo{journal}{LOMI Prepint E3, Leningrad}  (\bibinfo{year}{1979}).

\bibitem[{\citenamefont{Kivshar and Malomed}(1989)}]{KivsharMalomed1989}
\bibinfo{author}{\bibfnamefont{Y.~S.} \bibnamefont{Kivshar}} \bibnamefont{and}
  \bibinfo{author}{\bibfnamefont{B.~A.} \bibnamefont{Malomed}},
  \bibinfo{journal}{Rev. Modern Phys.} \textbf{\bibinfo{volume}{61}},
  \bibinfo{pages}{763} (\bibinfo{year}{1989}).

\bibitem[{\citenamefont{Otxoa et~al.}(2019)\citenamefont{Otxoa, Atxitia, Roy,
  and Chubykalo-Fesenko}}]{OtxoaAtxitiaRoyEtAl2019}
\bibinfo{author}{\bibfnamefont{R.~M.} \bibnamefont{Otxoa}},
  \bibinfo{author}{\bibfnamefont{U.}~\bibnamefont{Atxitia}},
  \bibinfo{author}{\bibfnamefont{P.~E.} \bibnamefont{Roy}}, \bibnamefont{and}
  \bibinfo{author}{\bibfnamefont{O.}~\bibnamefont{Chubykalo-Fesenko}},
  \bibinfo{journal}{arXiv preprint arXiv:1903.08034}  (\bibinfo{year}{2019}).

\bibitem[{\citenamefont{Barthem et~al.}(2013)\citenamefont{Barthem, Colin,
  Mayaffre, Julien, and Givord}}]{BarthemColinMayaffreEtAl2013}
\bibinfo{author}{\bibfnamefont{V.}~\bibnamefont{Barthem}},
  \bibinfo{author}{\bibfnamefont{C.}~\bibnamefont{Colin}},
  \bibinfo{author}{\bibfnamefont{H.}~\bibnamefont{Mayaffre}},
  \bibinfo{author}{\bibfnamefont{M.-H.} \bibnamefont{Julien}},
  \bibnamefont{and} \bibinfo{author}{\bibfnamefont{D.}~\bibnamefont{Givord}},
  \bibinfo{journal}{Nature Commun.} \textbf{\bibinfo{volume}{4}},
  \bibinfo{pages}{2892} (\bibinfo{year}{2013}).

\bibitem[{\citenamefont{Bar'yakhtar et~al.}(2006)\citenamefont{Bar'yakhtar,
  Chetkin, Ivanov, and Gadetskii}}]{bar2006dynamics}
\bibinfo{author}{\bibfnamefont{V.~G.} \bibnamefont{Bar'yakhtar}},
  \bibinfo{author}{\bibfnamefont{M.~V.} \bibnamefont{Chetkin}},
  \bibinfo{author}{\bibfnamefont{B.~A.} \bibnamefont{Ivanov}},
  \bibnamefont{and} \bibinfo{author}{\bibfnamefont{S.~N.}
  \bibnamefont{Gadetskii}}, \emph{\bibinfo{title}{Dynamics of Topological
  Magnetic Solitons: Experiment and Theory}}, vol. \bibinfo{volume}{129}
  (\bibinfo{publisher}{Springer}, \bibinfo{year}{2006}).

\bibitem[{\citenamefont{Thiaville and Nakatani}(2006)}]{ThiavilleNakatani2006}
\bibinfo{author}{\bibfnamefont{A.}~\bibnamefont{Thiaville}} \bibnamefont{and}
  \bibinfo{author}{\bibfnamefont{Y.}~\bibnamefont{Nakatani}},
  \bibinfo{journal}{Top. Appl. Phys.} \textbf{\bibinfo{volume}{101}},
  \bibinfo{pages}{161} (\bibinfo{year}{2006}).

\end{thebibliography}

\end{document}